\documentclass[%
 reprint,
 amsmath,amssymb,
 aps,
prl,
]{revtex4-1}

\usepackage{graphicx}% Include figure files
\usepackage{dcolumn}% Align table columns on decimal point
\usepackage{bm}% bold math
\begin{document}

%\preprint{APS/123-QED}

\title{Ultracold electrons via Near-Threshold Photoemission from Single-Crystal Cu(100)}% Force line breaks with \\

\author{Siddharth Karkare}
 \email{karkare@asu.edu}
 \affiliation{Physics Department, Arizona State University, Tempe AZ, USA 85282}%Lines break automatically or can be forced with \\

\author{Gowri Adhikari}%
\author{W. Andreas Schroeder}%
 \affiliation{%
 Department of Physics, University of Illinois at Chicago, Chicago, IL, USA 60607}%

\author{J. Kevin Nangoi}%
\author{Tomas Arias}%
\author{Jared Maxson}%
\affiliation{%
 Department of Physics, Cornell University, Ithaca, NY USA 14853}%

\author{Howard Padmore}
 \affiliation{Lawrence Berkeley National Lab, CA, USA 94720}

\date{\today}% It is always \today, today,

\begin{abstract}
Achieving a low mean transverse energy or temperature of electrons emitted from the photocathode-based electron sources is critical to the development of next-generation and compact X-ray Free Electron Lasers and Ultrafast Electron Diffraction, Spectroscopy and Microscopy experiments. In this paper, we demonstrate a record low mean transverse energy of 5 meV from the cryo-cooled (100) surface of copper using near-threshold photoemission. Further, we also show that the electron energy spread obtained from such a surface is less than 11.5 meV, making it the smallest energy spread electron source known to date: more than an order of magnitude smaller than any existing photoemission, field emission or thermionic emission based electron source. Our measurements also shed light on the physics of electron emission and show how the energy spread at few meV scale energies is limited by both the temperature and the vacuum density of states.  \end{abstract}

\maketitle

%\tableofcontents

The brightness of pulsed electron beams emitted from photoemission based sources (photocathodes) ultimately determines the performance of several applications like X-Ray Free Electron Lasers (XFELs) \cite{FEL1} and Ultrafast Electron Diffraction (UED) and microscopy experiments \cite{UED1}. For XFELs, a brighter electron beam would allow lasing at higher x-ray photon energies and with larger x-ray pulse energies. Brighter electron beams are also a key ingredient in the development of compact, university-scale XFELs \cite{DOE_report1}. For UED experiments, brighter electrons beams will allow the study of larger lattice sizes, macro-molecular assemblies and obtain information about the electronic structure along with the lattice structure \cite{DOE_report2}. Along with higher brightness, Ultrafast Electron Energy Loss Spectroscopy (U-EELS) techniques can tremendously benefit from a low energy spread of the electron source \cite{UEELS,UEELS1}. A lower energy spread from the electron source could allow for observation of vibrational modes of lattices at ultrafast time scales \cite{vibrational}.

For applications like stroboscopic UED (and microscopy) and U-EELS , which do not require more than a single to few electrons per pulse, the brightness is inversely proportional to both, the emission area on the photocathode (typically limited by diffraction limit of light to a few  $\mu$m) and the mean transverse energy (MTE). MTE is equivalent to the temperature of the electrons emitted in vacuum \cite{Advances}.  For applications like XFELs and single shot UED which require a large peak current density from the cathode the maximum possible brightness is directly proportional to the n$^\mathrm{th}$ power accelerating electric field (where n is between 1 and 2 depending on the application) and is inversely proportional to the MTE of the emitted electrons \cite{brightness_limit}. 
Thus, understanding and reducing the MTE or the equivalent temperature of electrons emitted from a photocathode is of paramount importance for obtaining brighter electron beams and improving the performance of all the aforementioned applications.

In practice, photocathodes used today for such applications are polycrystalline metals (typically Cu) or high-quantum-efficiency, low-electron-affinity semiconductors like alkali-antimonides (Cs$_3$Sb, K$_2$CsSb, Na$_2$KSb) or Cs$_2$Te \cite{Advances}. The MTE of electrons obtained from such photocathodes used today is a few 100 meV and is generally limited by the excess energy provided to the electrons above the work function by the incident photons. Based on the Sommerfield model of photoemission, Dowell and Schmerge showed the MTE is roughly equal to $E_{excess}/3$ where, $E_{excess}=\left(\hbar\omega-W\right)$ is the excess energy, defined as the difference between the photon energy $\hbar\omega$ the work function $W$ of the photocathode \cite{dowell}. Near the photoemission threshold, when the excess energy is close to or less than zero, the emission occurs from the tail of the Fermi distribution, limiting the MTE to $k_BT$, where $k_B$ is the Boltzmann constant and $T$ is the temperature of the electrons in the crystal \cite{vecchione}. For small laser fluences, when the electrons are in equilibrium with the lattice, $T$ is the temperature of the lattice. Thus, at the photoemission threshold, at room temperature, the MTE is limited to 25 meV. This near threshold limit has been experimentally demonstrated from polycrystalline Sb by Feng et al \cite{Feng}.

Reducing the MTE below 25 meV is possible by cooling the cathode to cryogenic temperatures. MTE as low as 20 meV has been demonstrated from alkali-antimonides by cooling down to 90 K using liquid nitrogen and measuring at the photoemission threshold \cite{cryocathode}. In this case the MTE measured was significantly higher than $k_BT$ (7.5 meV at 90 K). This large MTE was attributed to the effect of surface non-uniformities like roughness and work function variations \cite{chem_rough}. In order to minimize the effects of such variations on MTE, it is critical to emit electrons from a single crystalline cathode with an atomically ordered surface \cite{phys_chem_rough}. 

The MTE from atomically ordered surfaces of single crystalline materials is determined by the band-structure and the conservation of transverse momentum and energy that holds during photoemission from such surfaces \cite{schroeder1}. In such cases, assuming the effects of many-body interactions between photons, electrons and phonons during excitation and the effects of phonon scattering during emission \cite{many_body} are negligible, the effect of band-structure on MTE can be modelled accurately using the one-step photoemisison model \cite{one_step}. This effect has been demonstrated experimentally using the surface state of the Ag(111) surface. In this case, a non-monotonically increasing behaviour of MTE with excess energy was observed as predicted by the one-step photoemission model \cite{single_crystal}. However, emission from the relatively large transverse momentum states in the Ag(111) surface state restricted the MTE to values higher than 25 meV. Thus, in order to reduce MTE to the smallest possible value, along with using an atomically-ordered-single-crystal surface, it is essential to ensure that the band structure does not allow emission from large transverse momentum states.

In this paper, we aim to reduce all the above effects of excess energy, surface non-uniformities and emission from large transverse momentum states in the band-structure to obtain the smallest possible MTE. We measure the total energy and transverse momentum distributions (or equivalently the 3-D momentum distributions)  from the ordered Cu(100) surface cooled to 35 K using liquid helium at several wavelengths close to the photoemission threshold. Our measurements show an MTE as low as 5 meV (a factor of 4 lower than the smallest MTE measured to-date and a factor of 20-100 smaller than the MTE typically used for various applications) can be achieved. Furthermore, the total energy distribution measurements show an energy spread of electrons as low as 11.5 meV FWHM from this surface. This energy spread is more than an order of magnitude smaller than the smallest energy spread electron source demonstrated to-date (including thermal and field emission sources) making our results of great consequence to the development of U-EELS techniques along with UED and XFELS.

For this work the atomically clean and ordered Cu(100) surface was prepared via performing repeated ion-bombarding and annealing cycles on a commercially-purchased, mirror-polished, single crystal Cu(100) sample. 1 keV Ar$^+$ ions were used for ion-bombarding while annealing was performed at 600$^{\circ}$C for 30 minutes in an ultra high vacuum (UHV) chamber with a base pressure in the low 10$^{-10}$ torr range. The ion-bombarding and annealing cycles were performed until a sharp low energy electron diffraction (LEED) pattern of the (100) surface was obtained and Auger Electron Spectroscopy (AES) showed the surface to be free of carbon and oxygen contaminants. The sample was then transferred in UHV into a time-of-flight (ToF) based energy analyzer capable of measuring 3-D electron energy distributions of milli-eV energy scale electrons \cite{ToF}.

The energy analyzer comprises of the sample and a delay-line-based ToF detector arranged in a parallel plate configuration separated by $\sim$ 4 cm. A sub-picosecond pulsed laser is focused onto the sample. The intensity of the laser is low enough that no more than one electron is emitted per pulse. The emitted electron is accelerated towards the detector by an accelerating voltage of a few volts. The ToF detector measures the transverse position of the electron striking the detector and the time of flight of the electron from the sample to the detector. These measurements can be used to infer the transverse and longitudinal momentum of the electron at the time of emission and, consequently, the total energy and transverse momentum distributions. Further details of this setup are given elsewhere \cite{ToF}.    

For this work, measurements were performed at 2 accelerating voltages between the sample and the detector: 8 V and 4 V. Obtaining the transverse momentum distributions does not require any information other than the transverse position of the electrons on the detector and the time of flight, both of which are directly measured by the detector. MTE was calculated from these transverse distributions. The MTE values obtained at 8V and 4 V accelerating voltages are identical within the experimental uncertainty indicating that the effect of stray fields on the MTE measurement is negligible.

\begin{figure*}
    \centering
    \includegraphics[width=\textwidth]{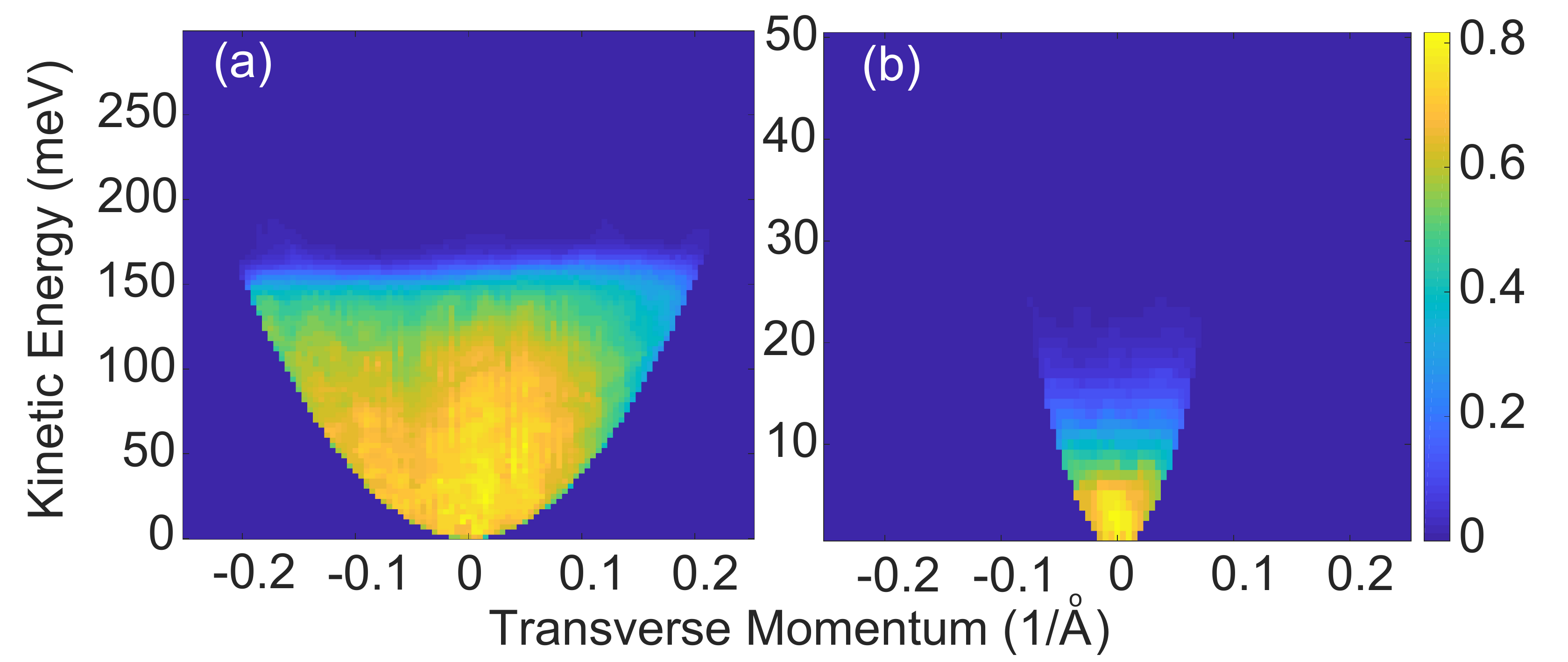}
\caption{Total kinetic energy vs transverse momentum distributions of emitted electrons using photon energy (a) 4.56 eV, (b) 4.43 eV. The transverse momentum spread with the 4.43 eV photon energy corresponds to 5 meV MTE. The figure shows the transverse momentum in only one transverse directions. The distributions are cylindrically symmetric in the transverse plane. The color bar is in arbitrary units.}
\end{figure*}

Obtaining the longitudinal momentum of the emitted electrons from the time of flight measurement, requires a detailed calibration of the work function difference and distance between the sample and the detector. 
The calibration was performed to ensure that the longitudinal momentum distribution does not change with the voltage applied between the sample and the detector and the energy of the Fermi edge in the total energy distribution increases with increasing photon energy. The details of the calibration procedure are given elsewhere \cite{ToF}. The calibration procedure gave a detector-sample distance of $40.3\pm0.1$ mm and the work function difference of $360\pm10$ meV. These values were used to obtain the longitudinal momentum distributions and the complete total-energy-transverse-momentum distributions. All the measurements were performed while the sample was cooled to 35 K using a continuous flow liquid helium cryostat connected to the sample holder.

A $\sim$150 fs pulse-width, wavelength-tunable, and frequency-tripled Ti-Sapphire oscillator with a repetition rate of 76 MHz provided the 4.2-4.9 eV UV photon energies used for the presented measurements. An acousto-optic pulse picker to was used to decrease the repetition rate to 3.8 MHz to stay below the maximum trigger rate of the delay-line-detector.
To minimize the effects of the photon energy spread on the measurements, the spectral width of this tunable sub-picosecond UV radiation source was reduced to $\sim$1.5 meV using a diffraction grating based monochromator. The photon energy could be tuned with an accuracy of 15 meV.

Figure 1 shows the total-energy-transverse-momentum distributions obtained at two photon energies of 4.43 eV and 4.56 eV. The transverse energy distributions are nearly cylindrically symmetric as expected from the band-structure. The MTE can be obtained by taking the second moment of the transverse momentum distributions.  Figure 2a shows the measured MTE as a function of the photon energy. The smallest MTE of 5 meV was measured at the photon energy of 4.43 eV. Figure 2b shows the total energy distributions obtained at various photon energies. As expected, the width of the energy distributions reduces with photon energy. At near-threshold photon energy of 4.43 eV the FWHM energy spread is measured to be less than 11.5 meV.

\begin{figure*}
    \centering
    \includegraphics[width=\textwidth]{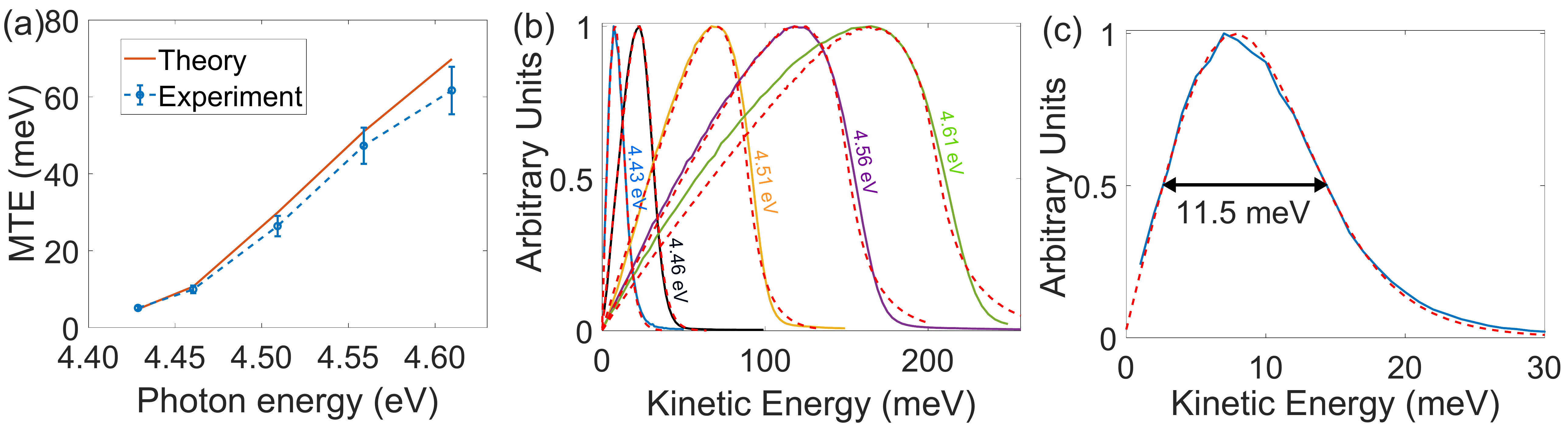}
\caption{(a) Measured and calculated values of MTE vs photon energy. MTE of 5 meV was measured at photon energy closest to the threshold of 4.42 eV. (b) Energy distribution curves at various photon energies. The solid curves are experimental measurement while the red dashed curves are theoretical calculations. (c) The 4.43 eV photon energy curve shows a FWHM spread of 11.5 meV}
\end{figure*}

Within the framework of the one-step model \cite{one_step}, photoemission can be modelled as a transition between an initial state in the crystal to a final state in vacuum under the influence of the perturbing Hamiltonian of incident light. The probability of the transition is proportional to:
\begin{equation}
    P=M^2f\left(E_i\right)\delta\left(k_{xi}-k_{xf}\right)\delta\left(k_{yi}-k_{yf}\right)\delta\left(E_f-E_i-\hbar\omega\right)
\end{equation}
where $E_f=\frac{\hbar^2}{2m_e}\left(k_{xf}^2+k_{yf}^2+k_{zf}^2\right)$ is the kinetic energy of the emitted electron; $k_{xf},k_{yf}$ and $k_{zf}$ are the wave-vectors of the electron emitted into vacuum ($z$ being the direction normal to the surface); $E_i$ is the initial energy of the electron inside the crystal and can be related to the initial wave-vectors $k_{xi},k_{yi}$ and $k_{zi}$ via the band-structure; $f\left(E_i\right)=1/\left(1+e^{\left(E_i-E_F\right)/k_BT}\right)$ is the Fermi distribution function and $E_F$ is the Fermi level and $M^2$ is the matrix element of the overlap integral between the wave-functions of the initial and the final state.    

The 3-D momentum distribution can be obtained by calculating the total probability of emission into a specific final state as
\begin{equation}
    N\left(k_{xf},k_{yf},k_{zf}\right)=\int\int\int Pdk_{xi}dk_{yi}dk_{zi}
\end{equation}

Calculating the overlap integrals to obtain the matrix element is fairly complex, however, inspired by the three step model of photoemission \cite{three_step}, in this work, we have assumed the matrix element to be $M^2=CT$, where $C$ is a constant and $T=\frac{4k_{zi}k_{zf}}{\left(k_{zi}+k_{zf}\right)^2}$ is the probability of transmission over the surface barrier.

The initial energy can be related to the initial wave-vectors via the band-structure. Figure 3 shows the band-structure of Cu along the high-symmetry paths in the Brillouin zone. The the $\Gamma-X$ direction is along the 100 direction. Due to the conservation of energy and transverse momentum, only electrons close to the Fermi level crossing in the $\Gamma-X$ direction can be emitted from the Cu(100) surface when the photon energies are within a few 100 meV of the threshold. The transverse momentum of electrons along the $\Gamma-X$ direction is zero. However, electrons emitted from bands that are transverse to the $\Gamma-X$ direction close to the Fermi crossing result in the non-zero MTE. There are no bulk states in the band-structure of Cu at 4-5 eV above this $\Gamma-X$ Fermi crossing. Therefore, the electrons are directly emitted into the vacuum states as shown in figure 3. 

Electrons near other Fermi level crossings or other locations on the Fermi surface, not close to the $\Gamma-X$ Fermi crossing, have too large of a transverse momentum to satisfy both the conservation of energy and transverse momentum simultaneously and hence cannot get emitted.

\begin{figure}
    \centering
    \includegraphics[width=0.5\textwidth]{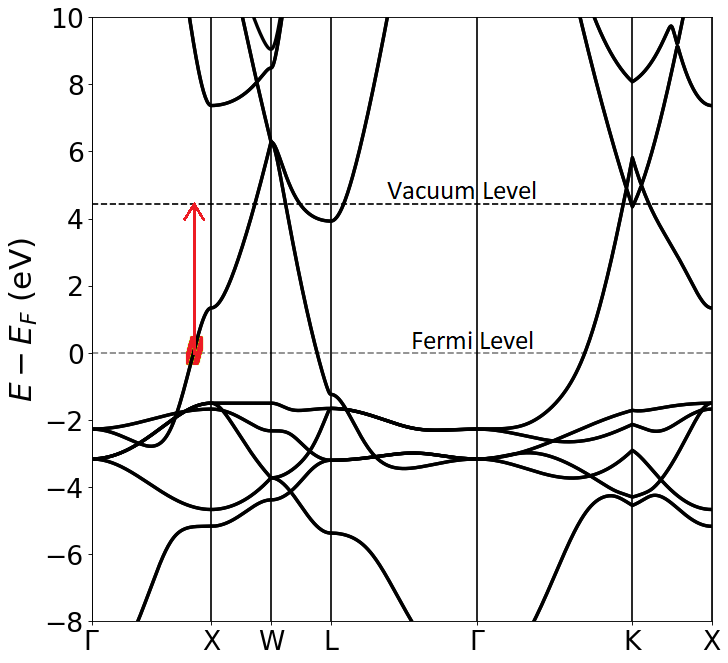}
\caption{Band structure of Cu along the high symmetry lines in the brillouin zone. Due to conservation of energy and transverse momentum, at photon energies close to the photoemission threshold, only electrons near the Fermi level crossing in the $\Gamma-X$ direction (region highlighted in red) can be emitted. There are no states in the band structure at the vacuum level at the wave-vector of this Fermi level crossing. Hence the excited electrons are emitted directly into vacuum. The red arrow indicates excitation directly into a vacuum state.}
\end{figure}

The band-structure of Cu around the $\Gamma-X$ Fermi level crossing can be given using the dispersion relation $E_i-E_F=\frac{1}{2}\frac{\hbar^2}{0.44m_e}\left(k_{xi}^2+k_{yi}^2\right)+\frac{1}{2}\frac{\hbar^2}{1.6m_e}k_{zi}^2-5.32~  \mathrm{eV}$. This relation was obtained via a 3-D quadratic fit to the band-structure calculated using density functional theory (DFT) \cite{schroeder3}. Note that the coefficients of the linear terms were found to be nearly zero and hence ignored. The coefficients of the quadratic cross terms are zero due to the crystal symmetry. Close to the $\Gamma-X$ Fermi crossing, the values of $k_{xi}$ and $k_{yi}$ are nearly zero, whereas the values of $k_zi$ are $\sim 1.5 \AA^{-1}$. The DFT calculations were performed using JDFTx, a plane-wave density-functional theory software \cite{jdftx}, with GGA-PBEsol exchange correlation functional \cite{pbesol} and GBRV ultrasoft pseudopotentials \cite{gbrv}. A plane-wave cutoff of 20 Hartree and a Brillouin zone sampling mesh of 12 x 12 x 12 were used. Using these parameters, the calculated optimum lattice constant is 3.57 $\AA$, within 1\% from the experimental value of 3.597 $\AA$ \cite{latt-const}.

The dispersion relation in the previous paragraph was used to calculate the 3-D momentum distribution $N\left(k_{xf},k_{yf},k_{zf}\right)$. The calculated distribution was convolved with a gaussian of a variable FWHM width equal to $0.3k_{zf}$ to account for the poor experimental resolution of the analyzer in the longitudinal direction at 4 V of accelerating voltage. The MTE and total energy distributions obtained from this theoretical 3-D momentum distribution are shown in figure 2 alongside the experimental data.
The experimental and theoretical data are in agreement within the experimental error. The calculations performed here assume a work function of 4.42 eV as no electrons could be detected below this photon energy.

The smallest MTE is 5 meV and is limited by the temperature, and the excess energy (due to the 15 meV inaccuracy in the laser wavelength tunability). At larger excess energies both the experimental and theoretical curves approximately follow the $E_{excess}/3$ trend proposed by Dowell and Schmerge \cite{dowell}.

As expected the total energy spread increases with the photon energy. The theoretically calculated energy distribution curves match the experimentally measured curves closely. The discrepancies at higher photon energies could be due to the approximate matrix elements used for the theoretical calculations. At the closest-to-threshold photon energy of 4.43 eV the FWHM in the energy distribution curve is 11.5 meV and is limited by the instrumental resolution, the temperature and the vacuum density of states.

The high energy edge in the distribution curves is given by the Fermi distribution, whereas the low-energy-side rise is due to the small density of vacuum states \cite{schroeder2} close to threshold and the low transmission probability as the kinetic energy (and hence $k_{zf}$) goes to zero. In the theoretical calculations the density of vacuum states, which increases as the square root of total kinetic energy, appears upon performing the summation over all possible momenta at a particular total kinetic energy, while the transmission probability is included via the matrix element. 
The energy dependent vacuum density of states, the energy dependent transmission probability and the Fermi distribution limit the total energy spread in the emitted electrons close to threshold. Thus, our measurements not only show a record low MTE and energy spread, but also shed light on several aspects of the physics of photoemission like the vacuum density of states and the vacuum transmission probability that are important near the threshold at very low kinetic energies.

The low 5 meV MTE from the Cu(100) surface implies a nearly two orders of magnitude brightness increase from the electron source for stroboscopic UED applications that require few to a single electron per pulse and the low energy spread can result in dramatically better energy resolution in U-EELS techniques. For larger charge density applications, the low quantum efficiency of the order of $10^{-8}$ for the near-threshold photon energy at 35 K will require the use of a large laser fluence. The MTE and hence the brightness will then be limited by non-linear photoemission effects of laser heating and multi-photon emission \cite{heating1, heating2, multi}. The laser fluence at which these nonlinear mechanisms cause a significant change in the MTE is still a matter under investigation. 

In order to mitigate the non-linear photoemission effects, use of higher quantum efficiency cathodes like alkali-antimonides will be essential. Our measurements show that obtaining MTE limited by the temperature is possible down to the few meV energy level, provided the effects of surface non-uniformities are minimized, making it critical to develop single crystal ordered surfaces of high quantum efficiency cathodes or with band structures and phonon interaction cross-sections that allow emission only from very low transverse momentum states even at high excess energies.

In this paper, by measuring the energy-momentum distributions of electrons photoemitted from the cryogenically cooled Cu(100) surface at the photoemission threshold, we have demonstrated a record low MTE and energy spread from the electron source. Our measurements shed light on various aspects of photoemission physics close to the threshold and show a way to obtain up to two orders of magnitude increased electron beam brightness for various ultrafast electron scattering and XFEL applications. For meV scale electron energies measured in this paper the de Broglie wavelength of the electrons is larger than a few nanometers. This new regime of low-energy-photoemission could in principle demonstrate new physics beyond the sudden approximation which is assumed in nearly all photoemission theories to date \cite{sudden_approximation}.

\begin{acknowledgements}
This work was supported by the U.S. National Science Foundation under Award No. PHY-1549132, the Center for Bright Beams and by the Director, Office of Science, Office of Basic Energy Sciences of the U.S. Department of Energy, under Contract KC0407-ALSJNT-I0013, DE-AC02-05CH11231 and DE-SC0017621.
\end{acknowledgements}
	
\end{document}